\documentclass{desyproc}

\begin{document}
\title{Elliptic Flow of Thermal Photons in Chemically Non-Equilibrated QCD Medium}

\author{{\slshape Akihiko Monnai$^1$}\\[1ex]
$^1$RIKEN BNL Research Center, Brookhaven National Laboratory, Upton, NY 11973, USA}

\contribID{87}

\confID{8648}  
\desyproc{DESY-PROC-2014-04}
\acronym{PANIC14} 
\doi  

\maketitle

\begin{abstract}
Heavy-ion experiments have recently revealed that azimuthal momentum anisotropy -- elliptic flow -- of direct photons from a QCD medium is a few times larger than hydrodynamic predictions. I present a possible explanation for the enhancement based on late quark chemical equilibration. The color glass theory indicates that the medium is initially gluon-rich. Thermal photons are then mainly produced after quarks are produced, at which point large flow anisotropy is developed. The numerical estimations indicate that slow chemical equilibration visibly enhances the elliptic flow of thermal photons. 
\end{abstract}

\section{Introduction}

The quark-gluon plasma (QGP) is a deconfined phase of quantum chromodynamics (QCD). After its experimental realization at BNL Relativistic Heavy Ion Collider (RHIC) \cite{RHIC:summary} and CERN Large Hadron Collider (LHC) \cite{Aamodt:2010pa}, there have been extensive studies on its properties. Large azimuthal momentum anisotropy, heavy quark diffusion and jet quenching observed in the experiments indicate that the quarks and gluons are strongly-coupled. The bulk medium follows the description of a relativistic fluid with small viscosity \cite{Schenke:2010rr}. 
On the contrary, photons and leptons are weakly coupled to the medium as they are not sensitive to the strong interaction. The transparency makes them unique probes to investigate the hot medium. Heavy particles, such as weak bosons, are mostly created at the time of collision while photons and dileptons can be produced thermally, which implies the existence of collectivity in those observables.

In this study, I focus on thermal photons. The photons created in the hard process are called prompt photons and those emitted softly from the medium are called thermal photons. One of the most useful observables for the quantification of the collectivity is elliptic flow $v_2$, which is defined in Fourier expansion of particle spectra with respect to the azimuthal angle
\begin{eqnarray}
\frac{dN}{d\phi} = \frac{N}{2\pi} [1 + 2v_1 \cos(\phi-\Psi_1) + 2v_2 \cos(\phi-\Psi_2) + 2v_3 \cos(\phi-\Psi_3) + ... ],
\end{eqnarray}
where $\phi$ is the angle in momentum space and $\Psi_n$ are the respective reaction planes. The quantity is sensitive to the magnitude of medium interaction as they are non-vanishing only when non-central geometry and/or fluctuation in the collision is converted into momentum anisotropy. Hadronic $v_2$ and $v_3$ are found large, which is considered as an evidence for the aforementioned strongly-coupled QGP. Surprisingly, on the other hand, direct photon $v_2$ is found large at RHIC \cite{Adare:2011zr} and LHC \cite{Lohner:2012ct}. It is a few times larger than hydrodynamic predictions and almost comparable to pion $v_2$, even though the quantity should reflect the small flow anisotropy in early stages of the time-evolution. This phenomenon is now recognized as ``photon $v_2$ puzzle". Recent experimental data also indicate the existence of large direct photon $v_3$ comparable to the pionic counterpart, which implies that the enhancement is at least partially due to the properties of the medium itself.

\section{Chemically non-equilibrated fluid}

A heavy-ion system before the collision is described as the color glass condensate, a state of saturated gluons \cite{McLerran:1993ni}. The system locally equilibrates in a short time $\sim 1$ fm/$c$ after the collision and form the quark-gluon plasma, which behaves as a fluid until it reaches freeze-out. An important observation is that thermalization and chemical equilibration do not necessarily occur simultaneously \cite{Monnai:2014xya}. 
When the system is chemically non-equilibrated at the onset of hydrodynamic evolution, thermal photon emission is suppressed because of the scarcity of quarks. This can lead to enhancement of photon $v_n$ as the contribution of later stages where the flow anisotropy is already large becomes effectively large \cite{Monnai:2014kqa}.

The inviscid hydrodynamic equations of motion of a chemically non-equilibrated system are energy-momentum conservation and rate equations for the energy momentum tensor and the number currents in the QGP phase:
\begin{eqnarray}
\partial_\mu T_q^{\mu \nu} + \partial_\mu T_g^{\mu \nu} &=& 0,  \\
\partial_\mu N_q^\mu &=& 2 r_b n_g - 2 r_b \frac{n_g^\mathrm{eq}}{(n_q^\mathrm{eq})^2} n_q^2 , \label{eq:dnqdt} \\
\partial_\mu N_g^\mu &=& (r_a - r_b) n_g - r_a \frac{1}{n_g^\mathrm{eq}} n_g^2 + r_b \frac{n_g^\mathrm{eq}}{(n_q^\mathrm{eq})^2} n_q^2 + r_c n_q - r_c \frac{1}{n_g^\mathrm{eq}} n_q n_g \label{eq:dngdt},
\end{eqnarray} 
where $r_a$, $r_b$ and $r_c$ are reaction rates for (a) gluon splitting, (b) the quark pair production and (c) gluon emission from a quark, respectively. The subscript {\em eq} denotes that the variable is in chemical equilibrium. Local thermalization of the quark and the gluon components is assumed and the total energy density and the total pressure are defined as $e = e_q + e_g$ and $P = P_q + P_g$. $n_q$ and $n_g$ are the number densities for quarks and gluons, respectively. Late quark chemical equilibration corresponds to $r_b < r_a, r_c$ because the pair creation is the process which changes the quark number density. Here the matter-antimatter degrees of freedom is included in $N_q^\mu$. It is note-worthy that the inverse of the reaction rate gives the chemical relaxation time $\tau_i$. Since the parton picture is no longer applicable in the hadronic phase, the system is simply assumed to be in chemical equilibrium below the crossover temperature. 

The hydrodynamic medium is characterized by the equation of state and the chemical reaction rates. The former is given by a hyperbolic interpolation of the parton gas of $u,d,g$ and the resonance gas of hadrons with mass up to $2.5$ GeV as $s = c(T) s_\mathrm{par} + [1-c(T)] s_\mathrm{had}$ where the interpolation function is $c(T) = \{1+ \tanh [(T-T_c)/\Delta T]\} /2$.
The connecting temperature $T_c = 0.17$ GeV and the crossover width $\Delta T = 0.017$ GeV are chosen. This preserves the consistency with the parton gas model used to estimate the equilibrium parton densities $n_q^\mathrm{eq}$ and $n_g^\mathrm{eq}$. The latter is parametrized as $r_i = c_i T$ $(i=a,b,c)$ where $c_i$ is the dimensionless parameter. The initial conditions for the energy density is based on an analytical Glauber model \cite{Kolb:2000sd}. The numerical estimation is performed in the (2+1)-dimensional boost invariant geometry. Thermal photon emission rate is based on the hyperbolic interpolation of the rates in the hadronic \cite{Traxler:1995kx} and the QGP phases \cite{Turbide:2003si}. 
Note that the parameters are chosen for demonstration of the chemical equilibration effects and precision analyses will be discussed elsewhere. 

\section{Numerical results}

\begin{figure}[hb]
\centerline{
\includegraphics[width=0.35\textwidth]{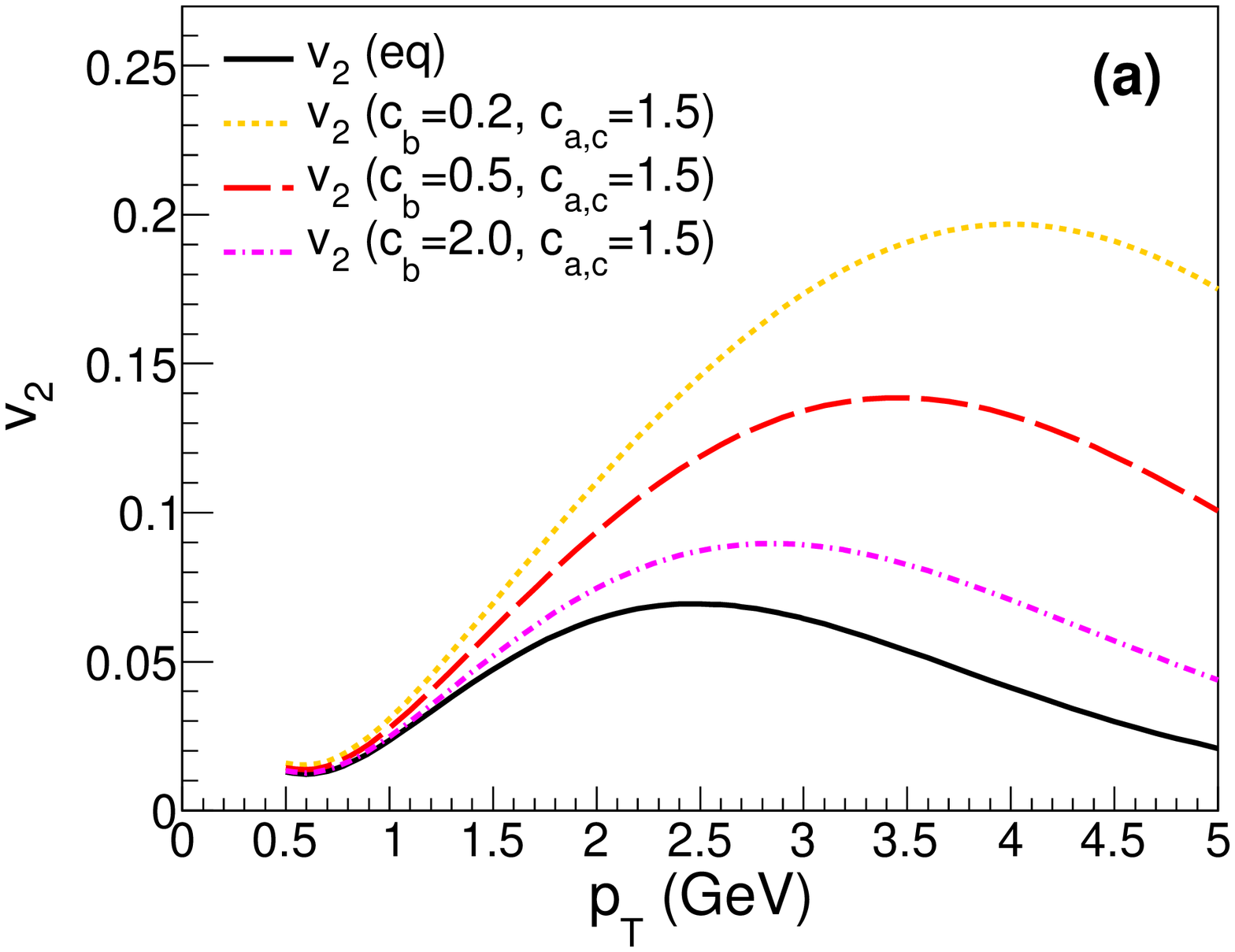}
\includegraphics[width=0.35\textwidth]{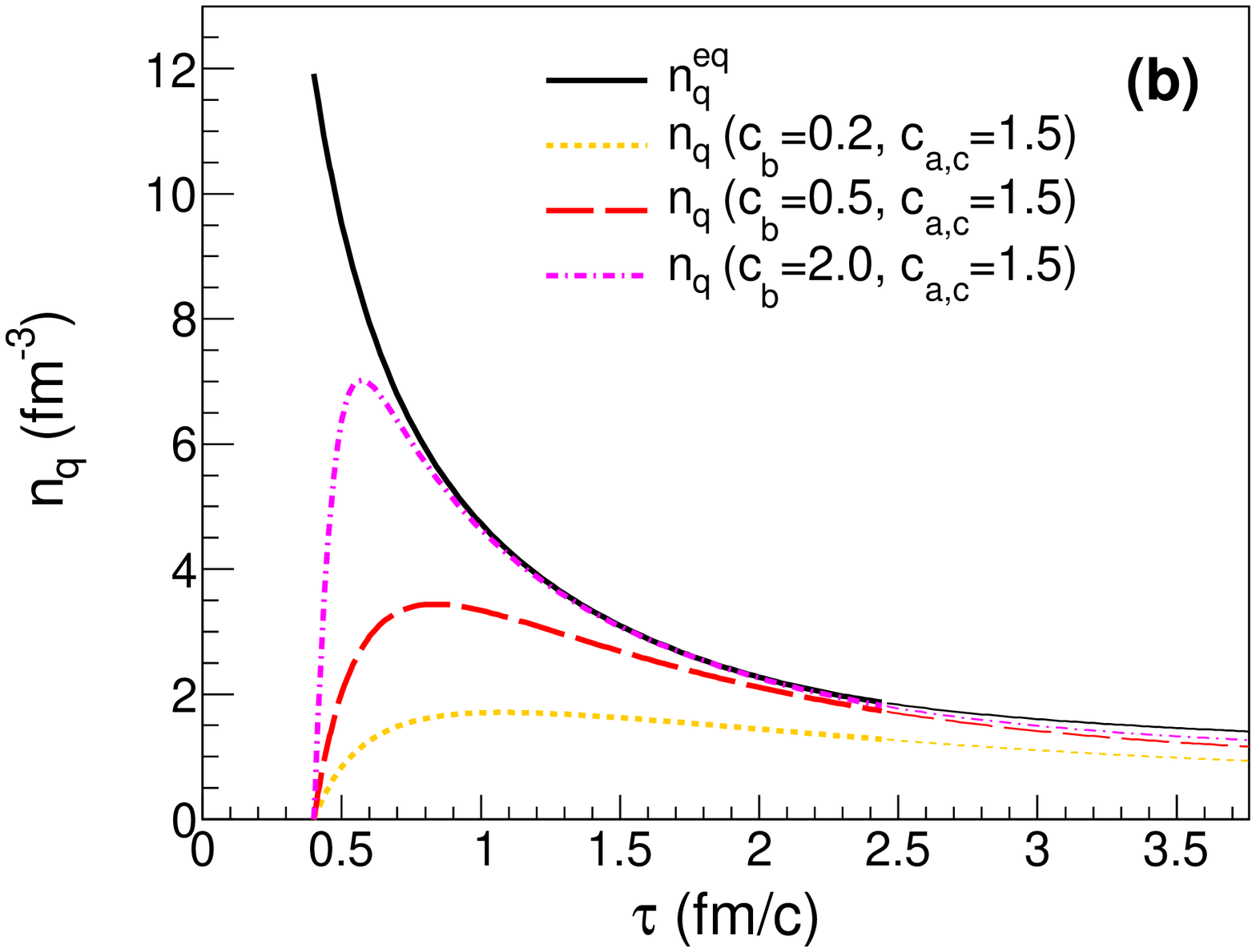}
}
\caption{(a) Elliptic flow of thermal photons and (b) time-evolution of the quark number density for different quark chemical equilibration rates. Thin lines for the number density indicate that the medium temperature is near the crossover $T < T_c + \Delta T$.}
\label{fig:amonnai1}
\end{figure}

Elliptic flow of thermal photons is shown in Fig.~\ref{fig:amonnai1} along with the time-evolution of the quark number density for different quark chemical equilibration parameters. $c_b = 0.2, 0.5$ and $2.0$ roughly corresponds to $\tau_\mathrm{chem} = 1/c_b T \sim 5.0, 2.0$ and $0.5$ fm/$c$ when the average medium temperature is $T \sim 0.2$ GeV. Au-Au collisions at $\sqrt{s_{NN}} = $200 GeV are considered. The impact parameter is set to $b=7$ fm at the thermalization time $\tau_\mathrm{th} = 0.4$ fm/$c$. One can see that late quark chemical equilibration leads to enhancement of thermal photon $v_2$. It is note-worthy that the typical chemical equilibration time suggested in a pre-equilibrium model \cite{Monnai:2014xya} is $2$~fm/$c$.

\begin{figure}[hb]
\centerline{
\includegraphics[width=0.35\textwidth]{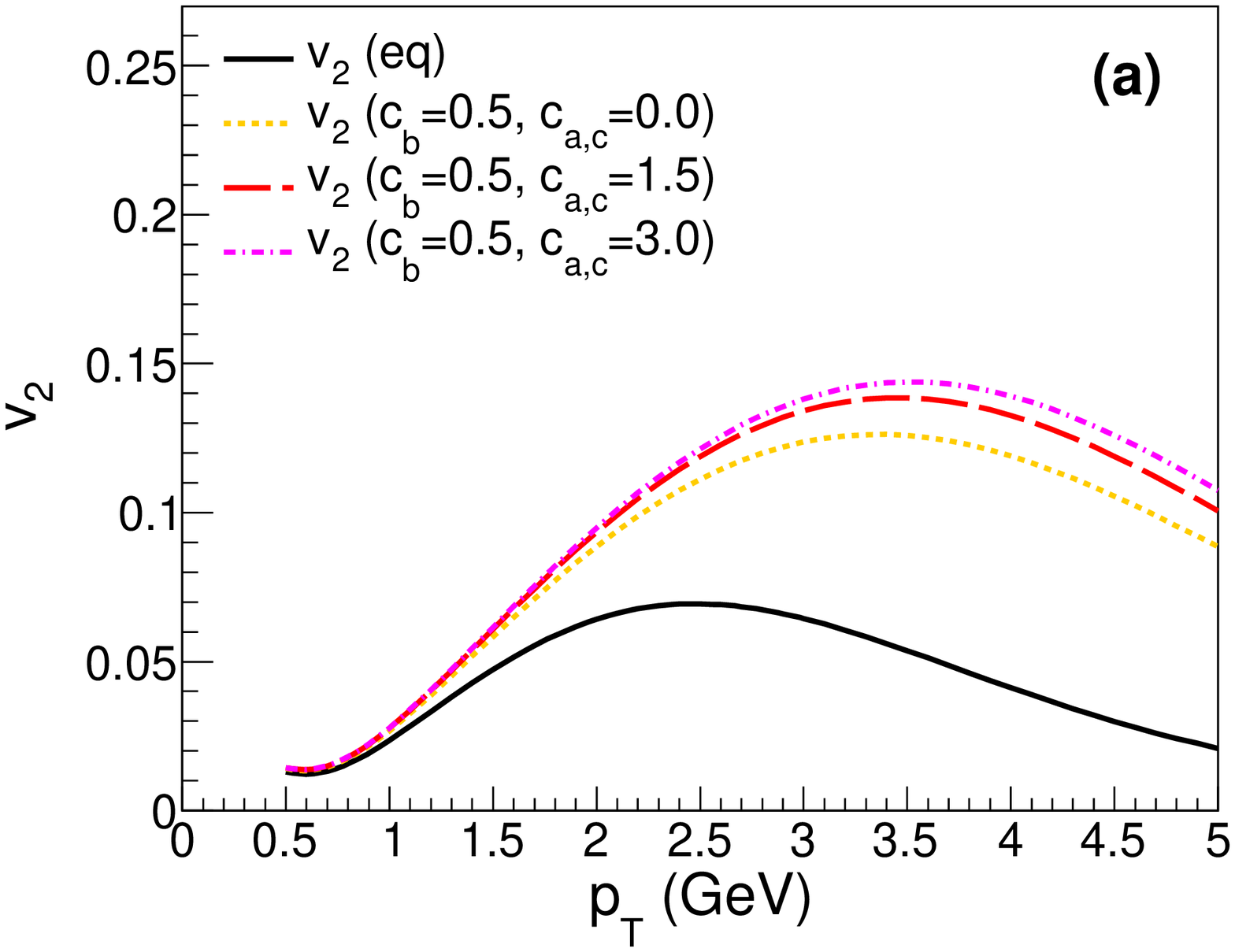}
\includegraphics[width=0.35\textwidth]{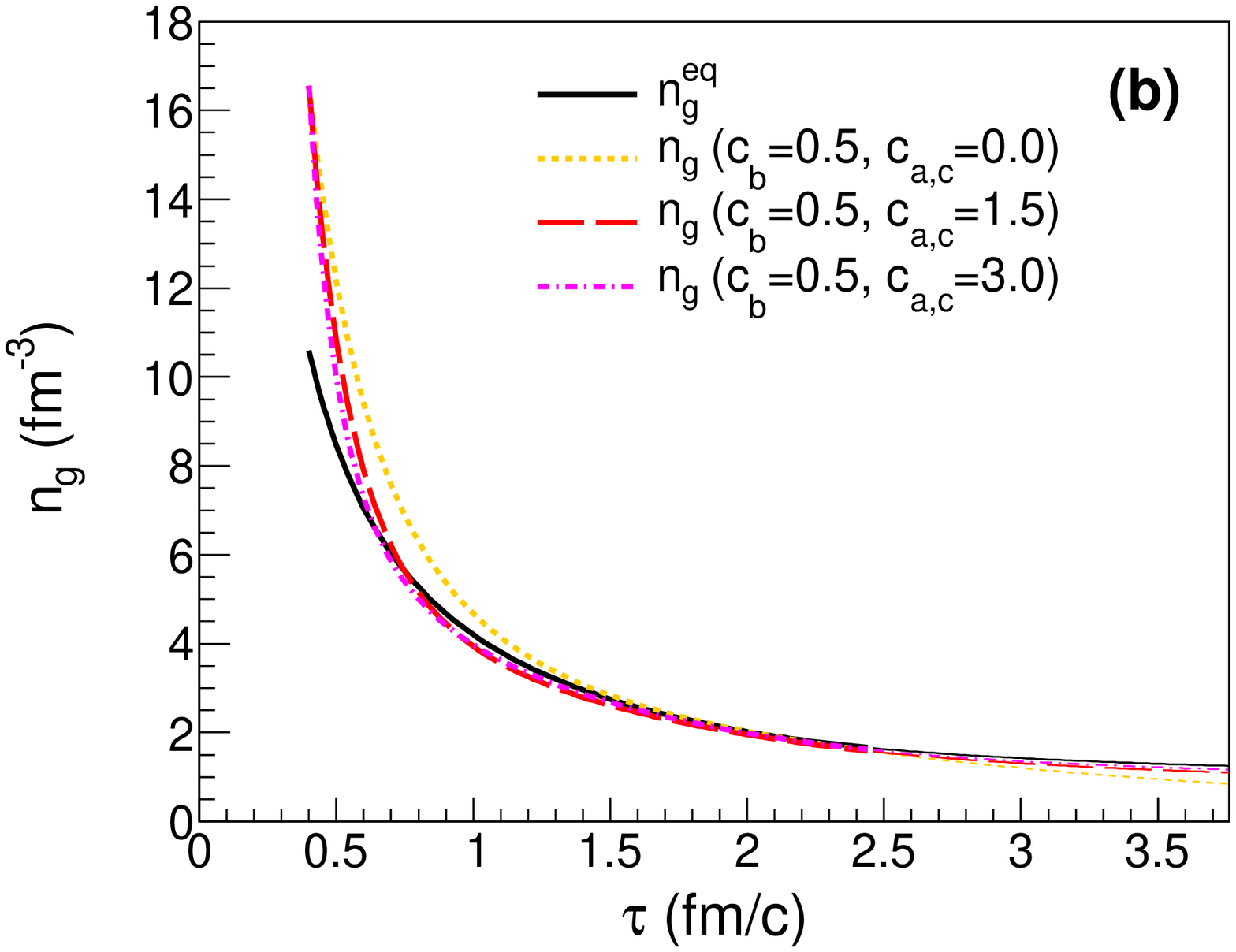}
}
\caption{(a) Elliptic flow of thermal photons and (b) time-evolution of the gluon number density for different gluon chemical equilibration rates. Thin lines for the number density indicate that the medium temperature is near the crossover $T < T_c + \Delta T$.}
\label{fig:amonnai2}
\end{figure}

Figure~\ref{fig:amonnai2} shows the dependence of thermal photon $v_2$ and that of the time-evolution of the gluon number density on gluon-involved equilibration parameters for a fixed $c_b$. The elliptic flow is slightly reduced for late gluon equilibration because gluons are initially overpopulated and larger number of gluons in the early stage leads to earlier production of quarks with small anisotropy. Quantitatively, it can be seen that the quark equilibration process plays bigger role in the thermal photon $v_2$ modification than the gluon one.

Transverse-momentum spectra of thermal photons show slight reduction for the late chemical equilibration cases due to the suppression in the photon emission. The magnitude is found to be not significant for the current parameter settings, but the introduction of more sophisticated photon emission rate and the dynamical equation of state would be important for quantitative discussion as the spectra and $v_2$ would both be sensitive to the input  \cite{Gelis:2004ep}.


\section{Conclusion and outlook}
Thermal photon elliptic flow from a chemically non-equilibrated QGP fluid is estimated by the newly-developed hydrodynamic model coupled to the rate equations for the parton number densities. This is motivated by the fact that a high-energy heavy-ion system is initially gluon-rich because the colliding nuclei are described as the color glass condensate. The numerical analyses indicate that thermal photon $v_2$ is visibly enhanced by late quark chemical equilibration, contributing positively to the resolution of the ``photon $v_2$ puzzle". The fact that the difference in thermalization and chemical equilibration times affects phenomenology implies that the interplay of pre- and post-equilibrium physics is important, which is in good analogy to the fact that thermal and chemical freeze-out have to be treated separately at the end of a hydrodynamic estimation for hadronic spectra. Future prospects include introduction of more quantitative photon emission rate, initial conditions, equation of state and transport processes as well as estimation of the contribution of prompt photons, which are naively believed to have almost vanishing anisotropy, for the comparison to the experimental data of direct photons. It would also be important to investigate the effects of other mechanisms, \textit{e.g.}, Ref.~\cite{Monnai:2014taa} for the comprehensive understanding of photon physics in heavy-ion collisions. 

\section*{Acknowledgments}

The work is inspired by fruitful discussion with B. M\"{u}ller. The author would like to thank for valuable comments by Y. Akiba and L. McLerran on the paper. The work of A.M. is supported by RIKEN Special Postdoctoral Researcher program.
 

\begin{footnotesize}



%

\end{footnotesize}


\end{document}